\documentclass[prd,aps,a4paper,superscriptaddress,twocolumn,nofootinbib]{revtex4}
\usepackage{graphicx}
\usepackage{color}
\usepackage{dcolumn}
\usepackage{bm}
\usepackage{slashed}
\usepackage{amsmath}
\usepackage{amsthm}
\usepackage{latexsym}
\usepackage{amssymb}
\usepackage{mathrsfs}
\usepackage{amsfonts}
\usepackage{url}
\usepackage{graphicx}
\usepackage{hyperref}

\newtheorem{theorem}{Theorem}[section]
\newtheorem{proposition}[theorem]{Proposition}
\allowdisplaybreaks
\begin{document}

\title{Constructing the canonical harmonic coordinates of Kerr metric to the fourth post-Minkowskian order}

\author{Zizheng Xing}
\email[Zizheng Xing:~]{Xingzz@mail.bnu.edu.cn}
\affiliation{School of Physics and Astronomy, Beijing Normal University, Beijing 100875, China}
\author{Xiaokai He}
\email[Xiaokai He:~]{sjyhexiaokai@hnfnu.edu.cn}
\affiliation{School of Mathematics and Statistics, and Hunan Provincial University Key Laboratory for Big Data Analysis and Application, Hunan First Normal University, Changsha 410205, China}
\author{Zhoujian Cao\footnote{corresponding author}}
\email[Zhoujian Cao:~]{zjcao@bnu.edu.cn}
\affiliation{School of Physics and Astronomy, Beijing Normal University, Beijing 100875, China}
\affiliation{School of Fundamental Physics and Mathematical Sciences, Hangzhou Institute for Advanced Study, University of Chinese Academy of Sciences, 
Hangzhou 310024, China}

\begin{abstract}
In this paper we construct the canonical harmonic coordinates of the Kerr metric within the multipolar post-Minkowskian (MPM) formalism to the fourth post-Minkowskian (4PM) order. Based on the well known Geroch--Hansen moments of Kerr metric and G\"ursel's theorem, we derive the exact canonical MPM moments $\mathrm{M}_L,\mathrm{S}_L$, which are free of any gauge moments. With these moments, we iteratively compute the gothic metric perturbation $h^{\mu\nu}_{\mathrm{can}}$ up to 4PM order and compute the 4PM canonical metric $g_{\mu\nu}^{\mathrm{can}}$. The resulting spatial and time components of these metrics are even functions of the spin parameter $a$ while the mixed components are odd. This parity property distinguishes the canonical coordinates from other harmonic coordinates. To contrast this minimal-gauge construction, we also extract the 1PM source moments of the Kerr metric in the Jiang--Lin coordinates. We find that the Jiang--Lin representation possesses non-vanishing gauge moments starting from the 1PM order, whereas in the canonical representation gauge moments vanish to all orders. This comparison highlights the canonical coordinates as the most gauge-pure representation of the Kerr metric in the MPM framework. The complete canonical metric for the Schwarzschild case is also computed to all PM orders.
A recent independent construction by Damgaard et al. using momentum-space recursion yields 4PM equivalent results expressed as a power series in $a$, providing a cross-validation of our closed-form 4PM canonical metric.
The coordinate transformation linking the canonical Kerr coordinates to previously known harmonic Kerr coordinates remains an open problem.
\end{abstract}

\maketitle

\section{Introduction}

The multipolar post-Minkowskian (MPM) formalism~\cite{Blanchet.2024} provides a systematic perturbative framework for constructing vacuum solutions of general relativity outside isolated sources.
In this formalism, the gravitational field is described by the gothic metric perturbation
$h^{\mu\nu} \equiv \sqrt{-g}\,g^{\mu\nu}-\eta^{\mu\nu}$, which is expanded in powers of the gravitational constant $G$.
The vacuum Einstein equations are solved iteratively order by order, leading to a general solution parametrized by six sets of symmetric and trace-free (STF) source moments:
two physical moments $\mathrm{I}_L, \mathrm{J}_L$ (mass-type and current-type respectively) and four gauge moments $\mathrm{W}_L,\mathrm{X}_L,\mathrm{Y}_L,\mathrm{Z}_L$.
The system can also be described isometrically by only two canonical moments $\mathrm{M}_L,\mathrm{S}_L$ in canonical coordinates where gauge moments vanish.
The MPM formalism has been proven to be a powerful tool for both gravitational wave generation problems and the analytical understanding of the gravitational field of isolated systems.

In this paper, we focus on the Kerr metric \citep{Kerr.1963}, the most fundamental exact solution describing a rotating black hole.
Our goal is to construct its canonical harmonic representation within the MPM framework, i.e., the particular harmonic coordinates in which all MPM gauge moments vanish identically.
Here the term ``canonical'' is used in the strict MPM sense or Thorne sense: a metric is canonical if it is generated by the functional $h_{\mathrm{can}(n)}^{\alpha\beta}[\mathrm{M}_L,\mathrm{S}_L]$ and parametrized solely by the two canonical moments, without any gauge moments \citep{Thorne.1980,Blanchet.1986,Blanchet.1998}.
The canonical MPM representation is theoretically important because it isolates the physical multipole content from pure gauge degrees of freedom.

We determine the canonical moments $\mathrm{M}_L,\mathrm{S}_L$ of Kerr metric by combining the well known Geroch--Hansen moments \cite{Geroch.1970,Hansen.1974} with G\"ursel's theorem \cite{Gursel.1983}, which relates Thorne moments (same as canonical MPM moments) to Geroch--Hansen moments.
With canonical moments in hand, we carry out the canonical MPM iteration algorithm up to the fourth post-Minkowskian (4PM) order.

The resulting canonical metric exhibits a clean parity separation: the spatial components $h^{ij}_{\mathrm{can}},g^{ij}_{\mathrm{can}},g_{ij}^{\mathrm{can}}$ and time component are even under the transformation $a \to -a$, while the time-spatial mixed components are odd. This property is a distinctive feature that distinguishes the canonical harmonic coordinates from other known harmonic coordinates.

To further appreciate the special status of the canonical coordinates, we also extract the 1PM source moments of the Kerr metric in the Jiang--Lin harmonic coordinates~\cite{Jiang.2014.GRG}.
By expanding the Jiang--Lin metric in powers of $G$ and matching it to the general MPM formalism, we find that its physical moments $\mathrm{I}^0_L,\mathrm{J}^0_L$ coincide with the canonical ones, but it carries a non-zero gauge moment $\mathrm{Z}_L$ starting from the 1PM order.
Thus, the Jiang--Lin coordinates contain additional gauge redundancy, whereas the canonical coordinates eliminate it completely.
This comparison demonstrates that the canonical coordinates constructed here are, in a precise MPM sense, the minimal-gauge harmonic representation of the Kerr metric.

After the completion of the present work, we became aware of a series of papers by Damgaard et al.~\cite{Damgaard.2024,Damgaard.2026star,Damgaard.2026}, which developed a momentum-space recursion algorithm to construct the harmonic metric for stationary sources. 
In Ref.~\cite{Damgaard.2026}, the authors call this metric ``gauge terms absent'', which is referred to as ``canonical'' in the MPM framework.
The earlier paper~\cite{Damgaard.2026star} applied the recursive algorithm to obtain a 2PM metric for a stationary source with general multipoles, and discussed its role as a possible exterior metric for a stationary star. 
When specialized to the Kerr multipoles, this 2PM metric reproduces the low-order multipole structure of the Kerr solution.

The subsequent paper~\cite{Damgaard.2026} applied the momentum-space recursion further to 4PM order in the Kerr case and obtained a 4PM metric as a power series in the spin parameter $a$. 
Expanding our closed-form canonical metric in powers of $a$, we find exact agreement with their series coefficients. This agreement provides a highly nontrivial cross-validation of our results and suggests that their recursive algorithm captures the same underlying canonical representation.

Since the ``gauge terms absent'' or the ``canonical'' metric is only one particular choice in the infinite class of harmonic coordinates, it generally lacks certain terms that are present in other harmonic metrics, such as the 2PM odd terms of $a$ in the Kerr case. 
These terms can be generated by performing additional residual harmonic gauge transformations~\citep{Damgaard.2026star,Damgaard.2026}.
In the MPM language, such transformations correspond to non-zero gauge moments.

The paper is organized as follows.
Section~II briefly reviews the MPM formalism for vacuum solutions and formula for inverse Laplacian used in our construction.
Section~III derives the canonical MPM moments of Kerr metric from the Geroch--Hansen moments via G\"ursel's theorem.
Section~IV presents the iterative computation of the canonical Kerr metric up to the 4PM order and discusses its parity properties.
Section~V presents the iterative computation of the canonical Schwarzschild metric to the full PM order and gives the coordinate transformation between the canonical harmonic coordinates and the usual Schwarzschild coordinates.
Section~VI performs the source moment analysis of the Jiang--Lin harmonic coordinates and compares them with the canonical ones.
Section~VII summarizes the paper and discuss the 3PM coordinate transformation from Boyer--Lindquist coordinates to the canonical one found in the current paper.

Throughout the paper we explicitly retain $G$ and $c$.  $\eta^{\mu\nu}=\mathrm{diag}\{-1,1,1,1\}$ is the flat metric.

\section{MPM Formalism of vacuum solutions}
Using the gothic metric perturbation or the radiation field
$h^{\mu\nu}\equiv\sqrt{-g}\,g^{\mu\nu}-\eta^{\mu\nu}$ as the basic variable, we have
\begin{align}
    g^{\mu\nu}=\dfrac{\eta^{\mu\nu}+h^{\mu\nu}}{\sqrt{-g}},\quad g_{\mu\nu}\equiv\mathrm{Inv}(g^{\mu\nu}),
\end{align}
where the determinant $g\equiv\mathrm{det}g_{\mu\nu}$ can also be expressed as
\begin{align}
    g=\mathrm{det}(\eta^{\mu\nu}+h^{\mu\nu}).
\end{align}
Indices of $h^{\mu\nu}$ are raised and lowered with the flat metric,
${h^{\mu}}_{\nu}\equiv h^{\mu\rho}\eta_{\rho\nu},\,h_{\mu\nu}\equiv h^{\rho\sigma}\eta_{\rho\mu}\eta_{\sigma\nu}$.
Using the matrix perturbation theory one obtains
\begin{subequations}\label{gofh}
\begin{align}
    g^{\mu\nu}&=\Omega^{-1}(\eta^{\mu\nu}+h^{\mu\nu}),\\
    \Omega&=\exp\left(-\sum_{i=1}^{\infty}\dfrac{(-1)^i}{2i}\mathrm{tr}(h^i)\right),\\
    \mathrm{tr}(h^n)&={h^{a_0}}_{a_{1}}{h^{a_1}}_{a_{2}}\cdots{h^{a_{n-2}}}_{a_{n-1}}{h^{a_{n-1}}}_{a_{0}},\\
    g_{\mu\nu}&=\Omega\left(\eta_{\mu\nu}+\sum^{\infty}_{k=1}(-1)^k(h^k)_{\mu\nu}\right),\\
    (h^n)_{\mu\nu}&=h_{\mu a_{1}}{h^{a_1}}_{a_{2}}\cdots{h^{a_{n-2}}}_{a_{n-1}}{h^{a_{n-1}}}_{\nu}.
\end{align}
\end{subequations}
Thus $g^{\mu\nu}$, $g_{\mu\nu}$ and $\mathrm{det}g=-\Omega^2$ can all be written as power series of $h$.

Define the quantity $\Lambda_{\mathrm{harm}}^{\alpha\beta}(h)$ as \citep{Blanchet.2024}
\begin{align}
    \Lambda_{\mathrm{harm}}^{\alpha\beta}(h)&=
    - h^{\mu\nu} \partial_{\mu\nu} h^{\alpha\beta}
    + \partial_\mu h^{\alpha\nu} \partial_\nu h^{\beta\mu}\notag\\
    &\quad+ \frac{1}{2} g^{\alpha\beta} g_{\mu\nu} \partial_\lambda h^{\mu\tau} \partial_\tau h^{\nu\lambda}
    + g_{\mu\nu} g^{\lambda\tau} \partial_\lambda h^{\alpha\mu} \partial_\tau h^{\beta\nu}
    \notag\\
    &\quad
    - g^{\alpha\mu} g_{\nu\tau} \partial_\lambda h^{\beta\tau} \partial_\mu h^{\nu\lambda}
    - g^{\beta\mu} g_{\nu\tau} \partial_\lambda h^{\alpha\tau} \partial_\mu h^{\nu\lambda}
    \notag\\
    &\quad+ \frac{1}{8} \left( 2 g^{\alpha\mu} g^{\beta\nu} - g^{\alpha\beta} g^{\mu\nu} \right)\notag\\
    &\quad\times\left( 2 g_{\lambda\tau} g_{\epsilon\pi} - g_{\tau\epsilon} g_{\lambda\pi} \right)
    \partial_\mu h^{\lambda\pi} \partial_\nu h^{\tau\epsilon},\label{Lambdaharm}
\end{align}
which is a nonlinear expression of $h$ and $\partial h$. The Einstein field equation in the harmonic gauge is equivalent to the relaxed equations
\begin{align}
    \Box h^{\alpha\beta}&=\dfrac{16\pi G}{c^4} (-g)T^{\alpha\beta}+\Lambda^{\alpha\beta}_{\mathrm{harm}}(h),\\
    \partial_\beta h^{\alpha\beta}&=0,
\end{align}
where $T^{\alpha\beta}$ stands for the energy-momentum tensor of matter.

Expanding $h^{\mu\nu}=\sum_{n\ge1} G^n h^{\mu\nu}_{(n)}$, one can decompose $\Lambda_{\mathrm{harm}}^{\alpha\beta}(h)$ in powers of $G$, yielding the expressions
\begin{align}
    \Lambda_{\mathrm{harm}}^{\alpha\beta}(h)&=\Lambda_{\mathrm{harm}}^{\alpha\beta}\Bigl(\sum_{n=1}^{\infty} G^n h_{(n)}\Bigr)\notag\\
    &=\sum_{n=2}^{\infty} G^n \Lambda_{\mathrm{harm}(n)}^{\alpha\beta}(h_{(1)},\cdots,h_{(n-1)}).
\end{align}
That is to say $\Lambda_{\mathrm{harm}(1)}^{\alpha\beta}=0$ and $\Lambda_{\mathrm{harm}(2)}^{\alpha\beta}(h_{(1)})$ gives the first non-trivial contribution. At the mean time we can see each $h_{(n)}^{\alpha\beta}$ has dimension of  $G^{-n}$.
In \eqref{Lambdaharm} the inverse metric $g^{\mu\nu}$ and the covariant metric $g_{\mu\nu}$ always appear in pairs, so the factors $\Omega^{-1}$ and $\Omega$ always cancel, making it straightforward to obtain the exact expression for $\Lambda_{\mathrm{harm}(n)}^{\alpha\beta}$.

Ref.~\cite{Blanchet.2024} reviews the multipolar post-Minkowskian (MPM) solution $h_{\mathrm{MPM}}^{\mu\nu}=\sum_{n\ge1} G^n h_{(n)}^{\mu\nu}$ of the harmonic vacuum field equations.
The MPM solution form is given there. The first-order term reads
\begin{widetext}
\begin{subequations}\label{h(1)}
\begin{align}
    h_{(1)}^{\mu\nu}&[\mathrm{I}_L,\mathrm{J}_L,\mathrm{W}_L,\mathrm{X}_L,\mathrm{Y}_L,\mathrm{Z}_L](t,\bm{x})=h^{\mu\nu}_{\mathrm{can}(1)}[\mathrm{I}_L,\mathrm{J}_L](t,\bm{x})+
    \partial^\mu\varphi^\nu_1 + \partial^\nu\varphi^\mu_1 - \eta^{\mu\nu}
    \partial_\lambda\varphi^\lambda_1,\\
    h^{00}_{\mathrm{can}(1)} &= -\dfrac{4}{c^2}\sum_{l\geq 0}\dfrac{(-1)^l}{l!}\partial_L \left( \dfrac{1}{r} \mathrm{I}_L (u)\right),\\
    h^{0i}_{\mathrm{can}(1)} &= \dfrac{4}{c^3}\sum_{l\geq 1}\dfrac{(-1)^l}{l!} \left\{ \partial_{L-1} \left( \dfrac{1}{r} \dot {\mathrm{I}}_{iL-1} (u)\right) + \dfrac{l}{l+1} \varepsilon_{iab} \partial_{aL-1} \left( \dfrac{1}{r} {\mathrm{J}}_{bL-1} (u)\right)\right\}, \\
    h^{ij}_{\mathrm{can}(1)}&=-\dfrac{4}{c^4}\sum_{l\geq 2}\dfrac{(-1)^l}{l!}\left\{\partial_{L-2} \left( \dfrac{1}{r}\ddot {\mathrm{I}}_{ijL-2} (u)\right) + \dfrac{2l}{l+1} \partial_{aL-2} \left( \dfrac{1}{r}\varepsilon_{ab(i} \dot {\mathrm{J}}_{j)bL-2}(u)\right)\right\},\\
    \varphi^0_1 &=\dfrac{4}{c^3}\sum_{l\geq 0} \dfrac{(-1)^l}{l!}\partial_L \left( \dfrac{1}{r} \mathrm{W}_L (u)\right),\\
    \varphi^i_1 &= -\dfrac{4}{c^4}\sum_{l\geq 0} \dfrac{(-1)^l}{l!}\partial_{iL} \left( {\dfrac{1}{r}}\mathrm{X}_L(u) \right)-\dfrac{4}{c^4}\sum_{l\geq 1} \dfrac{(-1)^l}{l!} \left\{ \partial_{L-1}\left( \dfrac{1}{r}{\mathrm{Y}}_{iL-1}(u) \right) + \dfrac{l \varepsilon_{iab}}{ l+1}\partial_{aL-1}\left( {\dfrac{1}{r}}{\mathrm{Z}}_{bL-1}(u) \right)\right\}.
\end{align}
\end{subequations}
\end{widetext}
Here $r=|\bm{x}|$ is the coordinate distance to the origin. We have closely followed the notation of \cite{Blanchet.2024} in the above equation. The functional form is parametrized by six sets of symmetric and trace-free (STF) source multipole moments $\{\mathrm{I}_L,\mathrm{J}_L,\mathrm{W}_L,\mathrm{X}_L,\mathrm{Y}_L,\mathrm{Z}_L\}$, where $\{\mathrm{I}_L,\mathrm{J}_L\}$ are referred to as the mass-type and the current-type source moments, and $\{\mathrm{W}_L,\mathrm{X}_L,\mathrm{Y}_L,\mathrm{Z}_L\}$ are referred to as the gauge source moments. The moments are also one-parameter functions of retarded time $u=t-|\boldsymbol{x}|/c$. The subscript $L$ is the shorthand of spatial indices $i_1i_2\cdots i_l$. The notation $A_{\left\langle L\right\rangle}$ denotes taking STF part of the tensor $A_L$. The monopole and dipole moments $\mathrm{I}$, $\mathrm{I}_i$ and $\mathrm{J}_i$ are constant in time, reflecting the conservation of mass, momentum and angular momentum.
For $n\ge2$, the functional $h_{(n)}^{\mu\nu}[\mathrm{I}_L,\mathrm{J}_L,\mathrm{W}_L,\mathrm{X}_L,\mathrm{Y}_L,\mathrm{Z}_L](t,\bm{x})$ is defined recursively by
\begin{subequations}\label{MPMalgorithm}
\begin{align}
    &h_{(n)}^{\mu\nu}:=u_{(n)}^{\mu\nu}+v_{(n)}^{\mu\nu},\\
    &u_{(n)}^{\mu\nu}:=\underset{B=0}{\mathrm{FP}}\Box_{\mathrm{ret}}^{-1}\Bigl[\Bigl(\frac{r}{r_0}\Bigr)^B \Lambda_{\mathrm{harm}(n)}^{\mu\nu}(h_{(1)},\cdots,h_{(n-1)})\Bigr],\\
    &(\Box_{\mathrm{ret}}^{-1}f)_{(t,\bm{x})}:=-\dfrac{1}{4\pi}\int \dfrac{\mathrm{d}^3\bm{y}}{|\bm{x-\bm{y}}|}f(t-|\bm{x-\bm{y}}|/c,\bm{y}),\\
    &v_{(n)}^{\mu\nu}:=\mathcal{V}^{\mu\nu}(w_{(n)}^\alpha),\quad w_{(n)}^{\alpha}=\partial_\beta u_{(n)}^{\alpha\beta}.
\end{align}
\end{subequations}
Here $r_0$ is an arbitrary constant length scale.
For brevity, we have omitted the functional parameters $[\mathrm{I}_L,\mathrm{J}_L,\mathrm{W}_L,\mathrm{X}_L,\mathrm{Y}_L,\mathrm{Z}_L]$ and the field arguments $(t,\bm{x})$ of the functional $h_{(n)}^{\mu\nu}$.
The symbol $\mathrm{FP}_{B=0}$ and the operator $\mathcal{V}^{\mu\nu}$ are defined in detail in the review~\cite{Blanchet.2024}.
The object $\mathrm{FP}_{B=0}\Box_{\mathrm{ret}}^{-1}[ (r/r_0)^B (\cdot) ]$ is called the generalized inverse d'Alembert operator, which should be distinguished from the ordinary inverse operator $\Box_{\mathrm{ret}}^{-1}$.
For a general function $f$, the integration in $\Box_{\mathrm{ret}}^{-1} f$ may not converge. In general, the functions encountered in the MPM algorithm are of this type, therefore the MPM algorithm employs the generalized inverse operator. If $\Box_{\mathrm{ret}}^{-1} f$ itself converges or is well-defined via analytic continuation, then the generalized inverse operator reduces to the ordinary operator
\begin{align}
    \Box_{\mathrm{ret}}^{-1}f=\underset{B=0}{\mathrm{FP}}\Box_{\mathrm{ret}}^{-1}\Bigl[\Bigl(\dfrac{r}{r_0}\Bigr)^B f\Bigr].
\end{align}

In the vacuum region, the MPM solution $h_{\mathrm{MPM}}^{\mu\nu}:=\sum_{n\ge1}G^n h_{(n)}^{\mu\nu}$, parametrized by the source moments, satisfies the relaxed harmonic vacuum Einstein field equations
\begin{subequations}
    \begin{align}
    \Box h^{\alpha\beta}_{\mathrm{MPM}}&=\Lambda_{\mathrm{harm}}^{\alpha\beta}(h_{\mathrm{MPM}}),\\
    \partial_\beta h^{\alpha\beta}_{\mathrm{MPM}}&=0.
\end{align}
\end{subequations}
The functional coefficients $h_{(n)}^{\mu\nu}$ satisfy, order by order
\begin{subequations}
    \begin{align}
    \Box h^{\alpha\beta}_{(n)}&=\Lambda_{\mathrm{harm}(n)}^{\alpha\beta}(h_{(1)},\cdots,h_{(n-1)}),\\
    \partial_\beta h^{\alpha\beta}_{(n)}&=0,
\end{align}
\end{subequations}

The MPM solution $h_{\mathrm{MPM}}^{\mu\nu}$ gives the most general solution of the harmonic vacuum Einstein field equations in the sense of \textbf{Theorem 2} in \cite{Blanchet.2024}, which can be restated as the following proposition.
\begin{proposition}
For any given integer $n$, for any given functions  $h_{[1]}^{\alpha\beta},h_{[2]}^{\alpha\beta},\cdots,h_{[n]}^{\alpha\beta}$ that are independent of $G$, past-stationary, vanishing at spatial infinity, and satisfying for all $1\le m\le n-1$
\begin{align}
    \partial_\beta h_{[m]}^{\alpha\beta}=0,\quad \Box h^{\alpha\beta}_{[m]}=\Lambda^{\alpha\beta}_{\mathrm{harm}(m)}(h_{[1]},\cdots,h_{[m-1]}),\label{eq1}
\end{align}
there exist six series of STF tensor
$\mathrm{A}_L^0,\mathrm{A}_L^1,\cdots,\mathrm{A}_L^{n-1}$ (independent of $G$) such that
\begin{align}\label{proposition}
\sum_{a=1}^n \left(G^a h_{(a)}^{\alpha\beta}\left[
    \mathrm{A}^0_L+\sum_{b=1}^{n-1} G^b\mathrm{A}^b_L\right]\right)
=\sum_{k=1}^n G^k h_{[k]}^{\alpha\beta}+O(G^{n+1}).
\end{align}
Here $f(G)=O(G^{n})$ denotes that
\begin{align}
    \exists\alpha, \forall G<\alpha, \exists \beta(\text{independent of }G),\quad |f(G)|<\beta G^{n}.
\end{align}
$G^{n}$ represents the upper bound of $f(G)$.
The symbol $\mathrm{A}$ stands collectively for the six source moments $\{\mathrm{I},\mathrm{J},\mathrm{W},\mathrm{X},\mathrm{Y},\mathrm{Z}\}$.
To cover the most general solution, the source moments are power series of $G$, for example $\mathrm{I}_L=\mathrm{I}_L^0+\sum_{b=1}^{n-1}G^b\mathrm{I}_L^b+O(G^n)$ is the MPM mass-type source moments.
\end{proposition}

Similarly the canonical MPM solution $h_{\mathrm{canMPM}}^{\mu\nu}[\mathrm{M}_L,\mathrm{S}_L]=\sum_{n\ge1}G^nh_{\mathrm{can}(n)}^{\mu\nu}[\mathrm{M}_L,\mathrm{S}_L]$ is also defined in \citep{Blanchet.2024}, which is parametrized by two sets of canonical moments $\mathrm{M}_L(u)$ and $\mathrm{S}_L(u)$. Functional $h^{\mu\nu}_{\mathrm{can}(1)}[\mathrm{M}_L,\mathrm{S}_L](t,\bm{x})$ is defined in (\ref{h(1)}b)-(\ref{h(1)}d) after replacing $\mathrm{I}_L,\mathrm{J}_L$ with $\mathrm{M}_L,\mathrm{S}_L$, and the higher-order functional $h_{\mathrm{can}(n)}^{\mu\nu}[\mathrm{M}_L,\mathrm{S}_L](t,\bm{x})$ is defined recursively similarly to \eqref{MPMalgorithm} after adding the subscript $_{\mathrm{can}}$. The canonical MPM solution $h_{\mathrm{canMPM}}^{\mu\nu}$ also satisfies the relaxed harmonic vacuum Einstein field equations and $h_{\mathrm{can}(n)}^{\mu\nu}$ also satisfies the equations order by order. Similarly the canonical moments $\mathrm{M}_L$ and $\mathrm{S}_L$ are power series of $G$.

If the canonical moments $\{\mathrm{M}_L(u_2),\mathrm{S}_L(u_2)\}$ and the source moments $\{\mathrm{I}_L(u_1),\mathrm{J}_L(u_1),\mathrm{W}_L(u_1),\mathrm{X}_L(u_1),\mathrm{Y}_L(u_1),\mathrm{Z}_L(u_1)\}$ satisfy the relations reviewed in \cite{Blanchet.2024}, the two metric functionals $g_{\mathrm{canMPM}}^{\alpha\beta}[\mathrm{M}_L,\mathrm{S}_L](t_2,\bm{x}_2)$ and $g_{\mathrm{MPM}}^{\mu\nu}[\mathrm{I}_L,\mathrm{J}_L,\mathrm{W}_L,\mathrm{X}_L,\mathrm{Y}_L,\mathrm{Z}_L](t_1,\bm{x}_1)$ are then mutually isometric under a coordinate transformation. In the isometric class sense the canonical MPM solution $h_{\mathrm{canMPM}}^{\mu\nu}$ gives the most general solution of the harmonic vacuum Einstein field equations.

For stationary configurations the inverse d'Alembert operator reduces to the inverse Laplacian operator
\begin{align}\label{LaplacianIntegration}
    (\Delta^{-1}f)_{(\bm{x})}:=-\dfrac{1}{4\pi}\int\dfrac{\mathrm{d}^3\bm{y}}{|\bm{x-\bm{y}}|}f(\bm{y}).
\end{align}
We have the formula
\begin{align}\label{inverseLaplacian}
    \Delta^{-1}\left(\dfrac{n_{\langle L \rangle}}{r^{\lambda}}\right)=\dfrac{n_{\langle L \rangle}}{(\lambda+l-2)(\lambda-l-3)r^{\lambda-2}}
\end{align}
provided the denominators do not vanish, where $n_{\langle L \rangle}$ is the shorthand of the tensor STF$(n_{i_1}n_{i_2}\cdots n_{i_l})$.

Strictly speaking, the integral \eqref{LaplacianIntegration} defining the left-hand side of \eqref{inverseLaplacian} converges absolutely only if $2<\lambda<3$. For almost all cases encountered in the MPM iteration, the integral does not converge, and formula \eqref{inverseLaplacian} already accomplishes a trivial regularization, namely a direct analytic continuation in the parameter $\lambda$ provided the denominators do not vanish.
All such trivial regularization operations are thereby packaged into the single algebraic formula \eqref{inverseLaplacian},  which may explain why our calculation is significantly simpler than that of Ref.~\cite{Damgaard.2026}.

If the denominator $\lambda-l-3$ vanishes, the finite-part operation $\mathrm{FP}_{B=0}$ is required, for example
\begin{align}\label{FPinverseLaplacian}
    \underset{B=0}{\mathrm{FP}}\Delta^{-1}\left(\dfrac{r^B}{r_0^B}\dfrac{n_{\langle L \rangle}}{r^{l+3}}\right)
    &=\frac{-(2l-1) \ln (r/r_0)-1}{(2 l-1)^2r^{l+1}}n_{\langle L \rangle}.
\end{align}

\section{Canonical MPM moments of Kerr}

For a vacuum, stationary and asymptotically flat spacetime,
Thorne \cite{Thorne.1980} introduced a class of coordinate systems, termed ACMC (asymptotically Cartesian and mass-centered to zero).
Within the ACMC class there exist infinitely many coordinate choices, and the metric components vary from one choice to another. Nevertheless, the inverse-distance expansion of the metric in any ACMC coordinate choice contains some invariant coefficients, the Thorne moments $\mathcal{G}_L$ and $\mathcal{S}_L$, which are independent of the ACMC coordinate choice.

For a stationary asymptotically flat spacetime with time-independent canonical moments $\mathrm{M}_L$ and $\mathrm{S}_L$, the metric $g_{\mathrm{canMPM}}^{\alpha\beta}$ and its coordinate system satisfy the requirement of ACMC. Therefore the Thorne moments $\mathcal{G}_L$ and $\mathcal{S}_L$ can be extracted from the expansion of $g_{\mathrm{canMPM}}^{\alpha\beta}$. The appendix of \cite{Gursel.1983} shows that the Thorne moments for such a metric $g_{\mathrm{canMPM}}^{\alpha\beta}[\mathrm{M}_L,\mathrm{S}_L]$ are actually the MPM canonical moments,
\begin{align}
    \mathcal{G}_L=\mathrm{M}_L,\quad \mathcal{G}_i=\mathrm{M}_i=0,\quad \mathcal{S}_L=\mathrm{S}_L.
\end{align}

In a geometric framework, Geroch and Hansen~\cite{Geroch.1970,Hansen.1974} defined another type of STF multipole moments $\mathfrak{m}_L,\mathfrak{s}_L$ for vacuum stationary spacetimes.
For an axisymmetric stationary spacetime with the axial direction $\bm{e}_z$, the coordinates-independent Geroch--Hansen scalar moments are also defined as \citep{Hansen.1974}
\begin{align}
    \mathfrak{m}_l=\dfrac{1}{l!}\mathfrak{m}_L(\bm{e}_z)^L,\quad \mathfrak{s}_l=\dfrac{1}{l!}\mathfrak{s}_L(\bm{e}_z)^L,
\end{align}
and the Geroch--Hansen tensor and scalar moments are related as~(see discussion in section 3 of~\cite{Hansen.1974})
\begin{align}
    \mathfrak{m}_L=(2l-1)!!\,\mathfrak{m}_l(\bm{e}_z)_{\langle L\rangle},\quad \mathfrak{s}_L=(2l-1)!!\,\mathfrak{s}_l(\bm{e}_z)_{\langle L\rangle}.
\end{align}
The Geroch--Hansen scalar moments of Kerr spacetime are given as \citep{Geroch.1970,Hansen.1974}
\begin{subequations}\label{KerrGHscalar}
\begin{align}
    \mathfrak{m}_l&=\left\{\begin{aligned}
        &(-1)^{l/2}\,M a^l,&&\text{$l$ even},\\
        &0,&&\text{$l$ odd},
    \end{aligned}\right.\\
    \mathfrak{s}_l&=\left\{\begin{aligned}
        &(-1)^{(l-1)/2}\,M c a^l,&&\text{$l$ odd},\\
        &0,&&\text{$l$ even}.
    \end{aligned}\right.
\end{align}
\end{subequations}
Thus the Geroch--Hansen tensor moments of Kerr spacetime are given by
\begin{subequations}\label{KerrGHtensor}
\begin{align}
    \mathfrak{m}_L&=\left\{\begin{aligned}
        &(-1)^{l/2}(2l-1)!!\,M a^l (\bm{e}_z)_{\langle L\rangle},&&\text{$l$ even},\\
        &0,&&\text{$l$ odd},
    \end{aligned}\right.\\
    \mathfrak{s}_L&=\left\{\begin{aligned}
        &(-1)^{(l-1)/2}(2l-1)!!\,M c a^l (\bm{e}_z)_{\langle L\rangle},&&\text{$l$ odd},\\
        &0,&&\text{$l$ even}.
    \end{aligned}\right.
\end{align}
\end{subequations}

Although Thorne moments and Geroch--Hansen moments are quite different in definition, G\"ursel~\cite{Gursel.1983} proved that for a vacuum, stationary and asymptotically flat metric in ACMC coordinates, the Thorne moments and the Geroch--Hansen moments are equivalent up to constant factors
\begin{align}
    \mathcal{G}_L=\dfrac{1}{(2l-1)!!}\mathfrak{m}_L,\quad
    \mathcal{S}_L=\dfrac{l+1}{2l(2l-1)!!}\mathfrak{s}_L.\label{theoremGursel.1983}
\end{align}
This theorem is also explained and generalized to the non-vacuum spacetime in \cite{Mayerson.2023}.

Since Throne moments and canonical MPM moments are same, we derive that the canonical MPM moments of Kerr metric must have the form
\begin{subequations}\label{MLSL}
\begin{align}
    \mathrm{M}_L&=\left\{\begin{aligned}
        &(-1)^{l/2}M a^l (\bm{e}_z)_{\langle L\rangle},&&\text{$l$ even},\\
        &0,&&\text{$l$ odd},
    \end{aligned}\right.\\
    \mathrm{S}_L&=\left\{\begin{aligned}
        &(-1)^{(l-1)/2}\dfrac{l+1}{2l}M c a^l (\bm{e}_z)_{\langle L\rangle},&&\text{$l$ odd},\\
        &0,&&\text{$l$ even}.
    \end{aligned}\right.
\end{align}
\end{subequations}

\section{4PM canonical Kerr metric}
We in this section compute the canonical Kerr metric iteratively to the fourth PM order, i.e., $\sum_{n=1}^4 G^n h_{\mathrm{can}(n)}^{\alpha\beta}[\mathrm{M}_L,\mathrm{S}_L]$.

\subsection{Calculation of the first PM order $h_{\mathrm{can}(1)}^{\alpha\beta}[\mathrm{M}_L,\mathrm{S}_L]$}
Since the canonical moments are constant in time, the first-order canonical metric $h_{\mathrm{can}(1)}^{\alpha\beta}$ reduces to
\begin{subequations}
\begin{align}
    h^{00}_{\mathrm{can}(1)} &= -\dfrac{4}{c^2}\sum_{l\geq 0}\dfrac{(-1)^l}{l!} \mathrm{M}_L\partial_L \Bigl( \dfrac{1}{r} \Bigr),\\
    h^{0i}_{\mathrm{can}(1)} &= \dfrac{4}{c^3}\sum_{l\geq 1}\dfrac{(-1)^l}{l!}\dfrac{l}{l+1}\varepsilon_{iab} {\mathrm{S}}_{bL-1}\partial_{aL-1} \Bigl( \dfrac{1}{r}\Bigr), \\
    h^{ij}_{\mathrm{can}(1)}&=0.
\end{align}
\end{subequations}
Here $r=|\bm{X}|=\sqrt{X^2+Y^2+Z^2}$.
Inserting \eqref{MLSL}, and noticing that
\begin{subequations}
\begin{align}
    (\bm{e}_z)_{\langle L\rangle}\partial_L\Bigl(\dfrac{1}{r}\Bigr)&=\dfrac{(-1)^l l!}{r^{l+1}}P_{l}(\mu),\\
    \varepsilon_{iab}(\bm{e}_z)_{\langle bL-1\rangle}\partial_{aL-1}\Bigl(\dfrac{1}{r}\Bigr)&=\dfrac{(-1)^{l-1} (l-1)!}{r^{l+1}}P_{l}'(\mu)(\bm{e}_z\times\bm{n})^i,
\end{align}
\end{subequations}
where $\bm{n}=\bm{X}/r,\mu=\bm{e}_z\cdot\bm{n}$, one derives
\begin{subequations}
\begin{align}
    h_{\mathrm{can}(1)}^{00}&=-\dfrac{4M}{c^2}\sum_{k\ge0}\dfrac{(-1)^k a^{2k}}{r^{2k+1}}P_{2k}(\mu),\\
    h_{\mathrm{can}(1)}^{01}
    &=+\dfrac{2aM}{c^2}\sum_{k\ge0}\dfrac{(-1)^k}{2k+1}\dfrac{a^{2k}}{r^{2k+2}}P_{2k+1}'(\mu)n_2,\\
    h_{\mathrm{can}(1)}^{02}
    &=-\dfrac{2aM}{c^2}\sum_{k\ge0}\dfrac{(-1)^k}{2k+1}\dfrac{a^{2k}}{r^{2k+2}}P_{2k+1}'(\mu)n_1.
\end{align}
\end{subequations}
The infinite summation can be formally rewritten in a closed form
\begin{subequations}\label{hcan(1)}
\begin{align}
    h_{\mathrm{can}(1)}^{00}&\overset{\mathrm{s}}{=}-\dfrac{4M R^3}{c^2(R^4+a^2Z^2)},\\
    h_{\mathrm{can}(1)}^{01}
    &\overset{\mathrm{s}}{=}+\dfrac{2aMR^3Y}{c^2(R^2 + a^2)(R^4 + a^2Z^2)},\\
    h_{\mathrm{can}(1)}^{02}
    &\overset{\mathrm{s}}{=}-\dfrac{2aMR^3X}{c^2(R^2 + a^2)(R^4 + a^2Z^2)},\\
    h^{03}_{\mathrm{can}(1)}&=h^{ij}_{\mathrm{can}(1)}=0.
\end{align}
\end{subequations}
The function $R(X,Y,Z)$ is defined implicitly by
\begin{align}
    R^4-R^2(X^2+Y^2+Z^2-a^2)-a^2Z^2=0,
\end{align}
or explicitly by
\begin{align}
    R=\sqrt{\dfrac{r^2-a^2+\sqrt{(r^2-a^2)^2+4a^2Z^2}}{2}}.
\end{align}
We use $A\overset{\mathrm{s}}{=}B$ to denote that $A$ equals the $r^{-1}$ series expansion of $B$. In other words, if only the series converge the first order PM metric equals the closed form shown in Eq.~(\ref{hcan(1)}).

\subsection{Iterate to the 4PM order $h_{\mathrm{can}(4)}^{\alpha\beta}$}

The MPM computation is straightforward and repetitive according to the algorithm Eq.~(\ref{MPMalgorithm}), and the computation is shown in the supplementary Mathematica notebook. In practice we start from the closed form of $h_{\mathrm{can}(1)}^{\alpha\beta}$ rather than from the infinite $r^{-1}$ series, which simplifies the handling of partial derivatives and products in $\Lambda_{\mathrm{harm}(n)}^{\alpha\beta}$.
During the iterative construction of the canonical Kerr metric to 4PM order, the non-trivial operations $\mathrm{FP}_{B=0}$ and $\mathcal{V}^{\alpha\beta}(w_{(n)})$ do not appear.
In each iteration step the inverse Laplacian $h_{\mathrm{can}(n)}^{\alpha\beta}=\Delta^{-1}\Lambda_{\mathrm{harm}(n)}^{\alpha\beta}$ is computed using the formula \eqref{inverseLaplacian}.

In practice, instead of directly applying the inverse Laplacian formula to each term, we first inspect the structure of $\Lambda_{\mathrm{harm}(n)}^{\alpha\beta}$ and guess a particular solution for $h^{\alpha\beta}_{\mathrm{can}(n)}$ using the method of undetermined functions, and then verify via the formula \eqref{inverseLaplacian} that this particular solution is indeed the required one $\Delta^{-1}\Lambda_{\mathrm{harm}(n)}^{\alpha\beta}$.

The 2PM solution $h_{\mathrm{can}(2)}^{\alpha\beta}$ is given by
\begin{subequations}
\begin{align}
    h_{\mathrm{can}(2)}^{00} &\overset{\mathrm{s}}{=} -\frac{M^2 R^2 \bigl(3 a^2+7 R^2\bigr)}{c^4\bigl(a^2+R^2\bigr) \bigl(a^2 Z^2+R^4\bigr)}, \\
    h_{\mathrm{can}(2)}^{01} &\overset{\mathrm{s}}{=} \frac{2 a M^2 R^2 Y}{c^4\bigl(a^2+R^2\bigr) \bigl(a^2 Z^2+R^4\bigr)}, \\
    h_{\mathrm{can}(2)}^{02} &\overset{\mathrm{s}}{=} -\frac{2 a M^2 R^2 X}{c^4\bigl(a^2+R^2\bigr) \bigl(a^2 Z^2+R^4\bigr)}, \\
    h_{\mathrm{can}(2)}^{03} &=0, \\
    h_{\mathrm{can}(2)}^{11}&\overset{\mathrm{s}}{=}-\frac{M^2 \bigl(R^2 X^2 \bigl(R^2-a^2\bigr)+a^2 \bigl(a^2+R^2\bigr) \bigl(R^2-Z^2\bigr)\bigr)}{c^4\bigl(a^2+R^2\bigr)^2 \bigl(a^2 Z^2+R^4\bigr)},\\
    h_{\mathrm{can}(2)}^{22}&\overset{\mathrm{s}}{=}-\frac{M^2 \bigl(R^2 Y^2 \bigl(R^2-a^2\bigr)+a^2 \bigl(a^2+R^2\bigr) \bigl(R^2-Z^2\bigr)\bigr)}{c^4\bigl(a^2+R^2\bigr)^2 \bigl(a^2 Z^2+R^4\bigr)},\\
    h_{\mathrm{can}(2)}^{33}&\overset{\mathrm{s}}{=}-\frac{M^2 Z^2}{c^4(a^2 Z^2+R^4)},\\
    h_{\mathrm{can}(2)}^{13}&\overset{\mathrm{s}}{=}\frac{-M^2 R^2 X Z}{c^4\left(a^2+R^2\right) \left(a^2 Z^2+R^4\right)},\\
    h_{\mathrm{can}(2)}^{23}&\overset{\mathrm{s}}{=}\frac{-M^2 R^2 YZ}{c^4\bigl(a^2+R^2\bigr) \bigl(a^2 Z^2+R^4\bigr)},\\
    h_{\mathrm{can}(2)}^{12}&\overset{\mathrm{s}}{=}\frac{-M^2 R^2 X Y \bigl(R^2-a^2\bigr)}{c^4\bigl(a^2+R^2\bigr)^2 \bigl(a^2 Z^2+R^4\bigr)}.
\end{align}
\end{subequations}
The 3PM solution $h_{\mathrm{can}(3)}^{\alpha\beta}$ is given by
\begin{subequations}
\begin{align}
    h_{\mathrm{can}(3)}^{00} &\overset{\mathrm{s}}{=} -\frac{8 M^{3} R^{3}}{c^{6}\bigl(a^2+R^2\bigr) \bigl(a^2 Z^2+R^4\bigr)}, \\
    h_{\mathrm{can}(3)}^{01} &\overset{\mathrm{s}}{=} \frac{2 a M^{3} R^{3} Y}{c^{6}\bigl(a^2+R^2\bigr)^{2} \bigl(a^2 Z^2+R^4\bigr)}, \\
    h_{\mathrm{can}(3)}^{02} &\overset{\mathrm{s}}{=} -\frac{2 a M^{3} R^{3} X}{c^{6}\bigl(a^2+R^2\bigr)^{2} \bigl(a^2 Z^2+R^4\bigr)},\\
    h_{\mathrm{can}(3)}^{03} &=h_{\mathrm{can}(3)}^{ij}=0.
\end{align}
\end{subequations}
The 4PM solution $h_{\mathrm{can}(4)}^{\alpha\beta}$ is given by
\begin{widetext}
\begin{subequations}
\begin{align}
    h_{\mathrm{can}(4)}^{00}&\overset{\mathrm{s}}{=}-\frac{M^4 \left(\left(a^2+R^2\right)^2 \arctan\left(\frac{a}{R}\right) \left(\left(a^2 Z^2+R^4\right) \arctan\left(\frac{a}{R}\right)-2 a R^3\right)+a^2 R^2 \left(16 a^4+a^2 \left(34 R^2-Z^2\right)+R^4\right)\right)}{4 a^4 c^8 \left(a^2+R^2\right)^2 \left(a^2 Z^2+R^4\right)},\\
    h_{\mathrm{can}(4)}^{01}&\overset{\mathrm{s}}{=}\frac{2 a M^4 R^2 Y}{c^8 \left(a^2+R^2\right)^2 \left(a^2 Z^2+R^4\right)},\\
    h_{\mathrm{can}(4)}^{02}&\overset{\mathrm{s}}{=}\frac{-2 a M^4 R^2 X}{c^8 \left(a^2+R^2\right)^2 \left(a^2 Z^2+R^4\right)},\\
    h_{\mathrm{can}(4)}^{03}&=0,\\
    h_{\mathrm{can}(4)}^{11}&\overset{\mathrm{s}}{=}\frac{M^4 \left(R^2 \left(X^2-Y^2\right) \left(a^3+R \left(a^2+R^2\right) \arctan\left(\frac{a}{R}\right)-a R^2\right)-a^3 \left(a^2+R^2\right) \left(R^2-Z^2\right)\right)}{2 a c^8 \left(a^2+R^2\right)^3 \left(a^2 Z^2+R^4\right)},\\
    h_{\mathrm{can}(4)}^{22}&\overset{\mathrm{s}}{=}-\frac{M^4 \left(R^2 \left(X^2-Y^2\right) \left(a^3+R \left(a^2+R^2\right) \arctan\left(\frac{a}{R}\right)-a R^2\right)+a^3 \left(a^2+R^2\right) \left(R^2-Z^2\right)\right)}{2 a c^8 \left(a^2+R^2\right)^3 \left(a^2 Z^2+R^4\right)},\\
    h_{\mathrm{can}(4)}^{33}&\overset{\mathrm{s}}{=}\frac{M^4 \left(\left(a^2+R^2\right) \arctan\left(\frac{a}{R}\right)-a R\right) \left(a^3 R \left(Z^2-2 R^2\right)+\left(a^2+R^2\right) \left(a^2 Z^2+R^4\right) \arctan\left(\frac{a}{R}\right)-a R^5\right)}{4 a^4 c^8 \left(a^2+R^2\right)^2 \left(a^2 Z^2+R^4\right)},\\
    h_{\mathrm{can}(4)}^{13}&\overset{\mathrm{s}}{=}\frac{M^4 R X Z \left(\left(a^2+R^2\right) \arctan\left(\frac{a}{R}\right)-a R\right)}{2 a c^8 \left(a^2+R^2\right)^2 \left(a^2 Z^2+R^4\right)},\\
    h_{\mathrm{can}(4)}^{23}&\overset{\mathrm{s}}{=}\frac{M^4 R Y Z \left(\left(a^2+R^2\right) \arctan\left(\frac{a}{R}\right)-a R\right)}{2 a c^8 \left(a^2+R^2\right)^2 \left(a^2 Z^2+R^4\right)},\\
    h_{\mathrm{can}(4)}^{12}&\overset{\mathrm{s}}{=}\frac{M^4 R^2 X Y \left(a^3+R \left(a^2+R^2\right) \arctan\left(\frac{a}{R}\right)-a R^2\right)}{a c^8 \left(a^2+R^2\right)^3 \left(a^2 Z^2+R^4\right)}.
\end{align}
\end{subequations}
\end{widetext}

From $\sum_{n=1}^4G^n h_{\mathrm{can}(n)}^{\alpha\beta}$, the 4PM metric $g^{\alpha\beta}_{\mathrm{can}}$ and $g_{\alpha\beta}^{\mathrm{can}}$ are also computed and shown in the supplementary Mathematica notebook.

Expanding our 4PM closed-form canonical metric in powers of $a$, we find exact agreement with the 4PM power series given in Damgaard et al.~\cite{Damgaard.2026} (see Eq.~(D.19,D.20) where $\mathfrak{h}’^{\mu\nu}=-h_{\mathrm{can}}^{\mu\nu}$).

In principle, one can compute the canonical MPM Kerr solution to any PM order.
In practice, however, the expression for $\Lambda_{\mathrm{harm}(5)}^{\alpha\beta}$ is far more complicated than $\Lambda_{\mathrm{harm}(4)}^{\alpha\beta}$ and contains terms such as $\arctan\left(a/R\right)$, which are absent at lower orders, making the evaluation of the inverse Laplacian $h_{\mathrm{can}(5)}^{\alpha\beta}$ extremely difficult.

It can be checked that the non-trivial operations $\mathrm{FP}_{B=0}$ and $\mathcal{V}^{\alpha\beta}(w_{(n)})$ are not required up to the 5PM order. This is because the expansions
\begin{subequations}
\begin{align}
    h_{\mathrm{can}(n)}^{\alpha\beta} &= \sum A^{\alpha\beta}_{(n),\lambda,L}\frac{n_{\langle L\rangle}}{r^{\lambda-2}},\\
    \Lambda_{\mathrm{harm}(n)}^{\alpha\beta} &= \sum (\lambda+l-2)(\lambda-l-3)A^{\alpha\beta}_{(n),\lambda,L}\frac{n_{\langle L\rangle}}{r^{\lambda}},
\end{align}
\end{subequations}
satisfy the following lower bounds on $\lambda-l-3$:
\begin{subequations}
\begin{align}
    n=1\quad \Rightarrow\quad \lambda-l-3&=0,\\
    n=2\quad \Rightarrow\quad \lambda-l-3&\ge -1,\quad  \lambda-l-3\neq 0,\\
    n=3\quad \Rightarrow\quad \lambda-l-3&\ge 2,\\
    n=4\quad \Rightarrow\quad \lambda-l-3&\ge 3,\\
    n=5\quad \Rightarrow\quad \lambda-l-3&\ge 4.
\end{align}
\end{subequations}
The 5PM bound $\lambda-l-3\ge 4$ has been verified by explicitly inspecting $\Lambda_{\mathrm{harm}(5)}^{\alpha\beta}$.
Since the denominator in the inverse Laplacian formula \eqref{inverseLaplacian} never vanishes for any $2\le n\le 5$, the ordinary inverse Laplacian is well defined at each step, and no finite-part operation $\mathrm{FP}_{B=0}$ is needed.

It is plausible that these two operations never appear at any PM order. This expectation is consistent with the natural requirement the canonical Kerr metric should not contain an arbitrary gauge constant $r_0$. (If the operation $\mathrm{FP}_{B=0}$ were required at some order, a logarithmic term $\ln(r/r_0)$ with an arbitrary length scale $r_0$ would be unavoidably introduced into the metric (see Eq.~\eqref{FPinverseLaplacian}, or the equation (89) in \cite{Blanchet.2024}).)

However, a rigorous proof of this property to all PM orders is currently beyond reach. In what follows, we adopt it as an assumption:
for the canonical Kerr solution, the ordinary inverse Laplacian $h_{\mathrm{can}(n)}^{\alpha\beta} = \Delta^{-1}\Lambda_{\mathrm{harm}(n)}^{\alpha\beta}$ is well-defined for all $n\ge 2$ through \eqref{inverseLaplacian}, without the finite-part operation $\mathrm{FP}_{B=0}$, and that the homogeneous correction $\mathcal{V}^{\alpha\beta}$ vanishes.

We also notice that for $n\le4$ the spatial components $h^{ij}_{\mathrm{can}(n)}$ and time component $h^{00}_{\mathrm{can}(n)}$ are even under $a \to -a$, while the mixed components $h^{0i}_{\mathrm{can}(n)}$ are odd. This parity property for canonical Kerr metric is very possibly valid for all PM orders. In the following we illustrate this property. 

For $n=1$ the parity property holds explicitly. We assume this parity property holds for all $m<n$ and then prove the parity property holds for $n$.

Denoting $h_1=\sum_{a=1}^{n-1}G^ah_{\mathrm{can}(a)}$, it follows from the induction hypothesis that $h_1^{00},h_1^{ij}$ are even functions of $a$, and $h_1^{0i}$ are odd functions of $a$.
The metric $g^{\mu\nu},g_{\mu\nu}$ are defined in \eqref{gofh}.
One can check that $\sqrt{-\mathrm{det}(\eta^{\mu\nu}+h^{\mu\nu})}$ is an even function of $a$, and therefore $g^{\alpha\beta}(h_1)$, $g_{\alpha\beta}(h_1)$ possess the same parity property with respect to $a$.
The $n$PM order source $\Lambda_{\mathrm{harm}(n)}^{\alpha\beta}$ is given by
\begin{align}
    \Lambda_{\mathrm{harm}(n)}^{\alpha\beta}&=\left[\Lambda_{\mathrm{harm}}^{\alpha\beta}\left(h_1\right)\right]_{\text{coefficient of }G^n}.
\end{align}
Since the metric is stationary, the time derivative vanishes in \eqref{Lambdaharm} and the Einstein summation involved in the derivative becomes spatial summation. 
One can easily check that for $\alpha\beta=00$ and $ij$, in each term the odd function $h^{0s}$ or $g^{0s}$ always appears in pairs, thus $\Lambda^{00}_{\mathrm{harm}}(h_1),\Lambda^{ij}_{\mathrm{harm}}(h_1)$ are even; 
for $\alpha\beta=0i$, in each term the odd function $h^{0s}$ or $g^{0s}$ always appears an odd number of times, thus $\Lambda^{0i}_{\mathrm{harm}}(h_1)$ is odd. 
Therefore $\Lambda^{\alpha\beta}_{\mathrm{harm}}(h_1)$ inherits the parity property.
Taking the $G^n$ coefficient, $\Lambda^{\alpha\beta}_{\mathrm{harm}(n)}$ inherits the parity property.
Since $h_{\mathrm{can}(n)}^{\alpha\beta}$ is the ordinary inverse Laplacian of $\Lambda_{\mathrm{harm}(n)}^{\alpha\beta}$, $h^{\alpha\beta}_{\mathrm{can}(n)}$ inherits the parity property.

In addition to the parity property with respect to $a\to -a$ discussed above, we observe another intriguing pattern in the canonical Kerr metric. The spatial components $h_{\mathrm{can}(n)}^{ij}$ vanish at every odd PM order. We have explicitly checked that
\begin{align}
h_{\mathrm{can}(1)}^{ij} = h_{\mathrm{can}(3)}^{ij} = h_{\mathrm{can}(5)}^{ij} = 0,
\end{align}
while $h_{\mathrm{can}(2)}^{ij}$ and $h_{\mathrm{can}(4)}^{ij}$ are non-vanishing.
This pattern indicates that the complete canonical spatial metric perturbation $h_{\mathrm{can}}^{ij} = \sum_{n=1}^{\infty} G^n h_{\mathrm{can}(n)}^{ij}$ is an even function of the mass parameter $m = GM/c^2$. Equivalently, only even powers of $G$ appear in the spatial sector of the canonical Kerr metric.
We are currently unable to prove this property to all PM orders. Nevertheless, its validity up to 5PM is firmly established by direct computation, and it would constitute a novel structural property of the canonical harmonic representation if confirmed in general.

\section{Canonical Schwarzschild metric}
Setting the spin parameter $a$ to zero, one obtains the canonical moments for the Schwarzschild spacetime
\begin{align}
    \mathrm{M}=M,\quad \mathrm{M}_L=0\;(l\ge1),\quad \mathrm{S}_L=0\;(l\ge1).
\end{align}
Only the mass-type canonical monopole is non-zero. The 4PM canonical Kerr metric then reduces to the 4PM Schwarzschild metric
\begin{subequations}\label{eq6}
\begin{align}
    h^{00}_{\mathrm{canS}(1)}&=-\dfrac{4M}{c^2r},\,
    h^{0i}_{\mathrm{canS}(1)}=0,\,
    h^{ij}_{\mathrm{canS}(1)}=0,\\[2pt]
    h^{00}_{\mathrm{canS}(2)}&=-\frac{7 M^2}{c^4 r^2},\,
    h^{0i}_{\mathrm{canS}(2)}=0,\,
    h^{ij}_{\mathrm{canS}(2)}=-\frac{M^2 n^in^j}{c^4 r^2},\\[2pt]
    h^{00}_{\mathrm{canS}(3)}&=-\frac{8 M^3}{c^6 r^3},\,
    h^{0i}_{\mathrm{canS}(3)}=0,\,
    h^{ij}_{\mathrm{canS}(3)}=0,\\[2pt]
    h^{00}_{\mathrm{canS}(4)}&=-\frac{8 M^4}{c^8 r^4},\,
    h^{0i}_{\mathrm{canS}(4)}=0,\,
    h^{ij}_{\mathrm{canS}(4)}=0.
\end{align}
\end{subequations}
Here $r=\sqrt{X^2+Y^2+Z^2}$, $\bm{n}=\bm{X}/r$, and the subscript ${\mathrm{S}}$ stands for the Schwarzschild case.

From Eq.~(\ref{eq6}) one can naturally propose the functions
\begin{subequations}
\begin{align}
    h^{00}_{\mathrm{canS}[n]}&=-\frac{8 M^n}{c^{2n} r^n},\,
    h^{0i}_{\mathrm{canS}[n]}=0,\,
    h^{ij}_{\mathrm{canS}[n]}=0,\quad (n\ge3)
\end{align}
\end{subequations}
to be the canonical solutions $h^{\alpha\beta}_{\mathrm{canS}(n)}$ to all PM orders. Here we use the notation $[n]$ to distinguish the trial functions from the actual solutions $h^{\alpha\beta}_{\mathrm{canS}(n)}$. We can prove they are indeed the solutions by induction. Using notation $m=G M/c^2$, we first compute the closed form of the formal sum $h^{\alpha\beta}_{\mathrm{can[S]}}=\sum_{a=1}^{\infty}G^a h^{\alpha\beta}_{\mathrm{canS}[a]}$
\begin{subequations}\label{hcanS}
\begin{align}
    h^{00}_{\mathrm{can[S]}}&\overset{\mathrm{s}}{=}1-\dfrac{(1+m/r)^3}{1-m/r},\\
    h^{0i}_{\mathrm{can[S]}}&=0,\\
    h^{ij}_{\mathrm{can[S]}}&=-\dfrac{m^2}{r^2}n^in^j,
\end{align}
\end{subequations}
and evaluate $\Lambda^{\alpha\beta}_{\mathrm{harm[S]}}=\Lambda^{\alpha\beta}_{\mathrm{harm}}(h_{\mathrm{can[S]}})$
\begin{subequations}\label{LambdaS}
\begin{align}
    \Lambda^{00}_{\mathrm{harm[S]}}&\overset{\mathrm{s}}{=}-\frac{2 m^2 \left(r+m\right) \left(m^2 -4 m r+7 r^{2}\right)}{r^{4} \left(r-m\right)^3},\\
    \Lambda^{0i}_{\mathrm{harm[S]}}&=0,\\
    \Lambda^{ij}_{\mathrm{harm[S]}}&=-\dfrac{2m^2}{r^4}\delta_{ij}+\dfrac{4m^2 }{r^4}n^in^j.
\end{align}
\end{subequations}

For $n\ge3$, assuming that $h^{\alpha\beta}_{\mathrm{canS}(m)}=h^{\alpha\beta}_{\mathrm{canS}[m]}$ holds for all $1\le m\le n-1$, we then compute the $n$-th order source $\Lambda_{\mathrm{harm}(n)}^{00}(h_{\mathrm{canS}(1)},\cdots,h_{\mathrm{canS}(n-1)})$
\begin{align}
    \Lambda_{\mathrm{harm}(n)}^{\alpha\beta}&=\Bigg[\Lambda_{\mathrm{harm}}^{\alpha\beta}\Bigg(
    \sum_{a=1}^{n-1} G^a
    \underset{{\text{coefficient of }G^n}}{h_{\mathrm{canS}(a)}\Bigg)\Bigg]}.
\end{align}
Since only the $G^n$ coefficient is required, terms of order $G^n$ and higher in the argument of $\Lambda_{\mathrm{harm}}^{\alpha\beta}$ may be added without affecting the result. Hence
\begin{align}
    \Lambda_{\mathrm{harm}(n)}^{\alpha\beta}&=\Bigg[\Lambda_{\mathrm{harm}}^{\alpha\beta}\Bigg(\sum_{a=1}^{n-1} G^a h_{\mathrm{canS}(a)}
    +\sum_{b=n}^{\infty}G^b
    \underset{{\text{coefficient of }G^n}}{h_{\mathrm{canS}[b]}\Bigg)\Bigg]}.
\end{align}
By the induction hypothesis, $\sum_{a=1}^{n-1} G^a h_{\mathrm{canS}(a)}+\sum_{b=n}^{\infty}G^b h_{\mathrm{canS}[b]}$ is precisely the sum in \eqref{hcanS}. Applying $\Lambda_{\mathrm{harm}}^{\alpha\beta}$ to this sum yields \eqref{LambdaS}. Extracting the $G^n$ coefficient gives
\begin{subequations}
\begin{align}
    \Lambda_{\mathrm{harm}(n)}^{00}(h_{\mathrm{canS}(1)},\cdots,h_{\mathrm{canS}(n-1)})
    &=-\dfrac{8n(n-1)M^n}{c^{2n}r^{n+2}},\\
    \Lambda_{\mathrm{harm}(n)}^{0i}(h_{\mathrm{canS}(1)},\cdots,h_{\mathrm{canS}(n-1)})
    &=0,\\
    \Lambda_{\mathrm{harm}(n)}^{ij}(h_{\mathrm{canS}(1)},\cdots,h_{\mathrm{canS}(n-1)})
    &=0\; (n\ge3).
\end{align}
\end{subequations}
Applying the generalized inverse d'Alembert operator, we obtain
\begin{align}
    u_{(n)}^{00}&=-\dfrac{8M^n}{c^{2n}r^n},\quad
    u_{(n)}^{0i}=0,\quad
    u_{(n)}^{ij}=0,\\
    w_{(n)}^\alpha&=\partial_\beta u_{(n)}^{\alpha\beta}=0,\quad
    v^{\alpha\beta}_{(n)}=\mathcal{V}^{\alpha\beta}(w_{(n)})=0,\\
    h_{\mathrm{canS}(n)}^{00}&=-\dfrac{8M^n}{c^{2n}r^n},\quad
    h_{\mathrm{canS}(n)}^{0i}=0,\quad
    h_{\mathrm{canS}(n)}^{ij}=0,
\end{align}
which confirms $h_{\mathrm{canS}(n)}^{\alpha\beta}=h_{\mathrm{canS}[n]}^{\alpha\beta}$ and closes the induction. Thus we have determined the complete canonical Schwarzschild solution $h_{\mathrm{canS}}^{\mu\nu}$ to all PM orders.

The metric corresponding to $h_{\mathrm{canS}}^{\alpha\beta}$ is
\begin{subequations}
\begin{align}
    g_{00}^{\mathrm{canS}}&\overset{\mathrm{s}}{=}-1+\dfrac{2m}{m+r},\\
    g_{0i}^{\mathrm{canS}}&=0,\\
    g_{ij}^{\mathrm{canS}}&\overset{\mathrm{s}}{=}\delta_{ij}-\dfrac{(m+r)m^2}{(m-r)r^2}n^in^j.
\end{align}
\end{subequations}
This canonical Schwarzschild coordinate system $\{cT,X,Y,Z\}$ is related to the usual Schwarzschild coordinates $\{ct,r_{\mathrm{S}},\theta,\varphi\}$ by
\begin{align}
&\left\{
\begin{aligned}
    T&=t,\\
    X&=(r_{\mathrm{S}}-m)\cos\varphi\,\sin\theta,\\
    Y&=(r_{\mathrm{S}}-m)\sin\varphi\,\sin\theta,\\
    Z&=(r_{\mathrm{S}}-m)\cos\theta.
\end{aligned}
\right.
\end{align}
It is worth noting that this canonical harmonic coordinate system for the Schwarzschild spacetime constructed above coincides with the well-known harmonic Schwarzschild coordinates (Eq. (8.2.15) of \cite{Weinberg.1972}). 
Using a momentum-space recursion, Damgaard et al.~\cite{Damgaard.2024} obtained the same harmonic Schwarzschild metric, providing an independent verification of our canonical construction in the Schwarzschild case.
Moreover, the canonical moments consist solely of the mass monopole $\mathrm{M}=M$, with all higher mass-type and current-type moments vanishing identically. This is consistent with the expectation that the canonical representation of a spherically symmetric spacetime contains only the minimal monopole content.


\section{Source moments of Kerr in Jiang--Lin coordinates and comparison to canonical harmonic coordinates}
The vacuum Kerr metric is usually expressed in Boyer--Lindquist coordinates $\{x^0=ct,r_{\mathrm{BL}},\theta,\varphi\}$ as \citep{Boyer.Lindquist.1967}
\begin{align}
    \mathrm{d}s_{\mathrm{BL}}^2&=-\Bigl(1-\dfrac{2mr_{\mathrm{BL}}}{r_{\mathrm{BL}}^2+a^2\cos^2\theta}\Bigr)(\mathrm{d}x^0)^2\notag\\
    &\quad+\dfrac{r_{\mathrm{BL}}^2+a^2\cos^2\theta}{r_{\mathrm{BL}}^2+a^2-2mr_{\mathrm{BL}}}\mathrm{d}r_{\mathrm{BL}}^2+(r_{\mathrm{BL}}^2+a^2\cos^2\theta)\mathrm{d}\theta^2\notag\\
    &\quad+\Bigl(r_{\mathrm{BL}}^2+a^2+\dfrac{2mr_{\mathrm{BL}}a^2\sin^2\theta}{r_{\mathrm{BL}}^2+a^2\cos^2\theta}\Bigr)\sin^2\theta \mathrm{d}\varphi^2\notag\\
    &\quad-\dfrac{4mr_{\mathrm{BL}}a\sin^2\theta}{r_{\mathrm{BL}}^2+a^2\cos^2\theta}\mathrm{d}x^0\mathrm{d}\varphi,\label{ds2BL}
\end{align}
where $m=GM/c^2$. The parameters $M$ and $a$ are independent, and both $m$ and $a$ have dimensions of length.
Jiang and Lin \cite{Jiang.2014.GRG,Jiang.2014.PRD} introduced a coordinate transformation
\begin{align}\label{Jacobi}
&\left\{
\begin{aligned}
T_{\mathrm{JL}}&=t,\\
X_{\mathrm{JL}}&=\sqrt{R_{\mathrm{JL}}^2+a^2}\cos\varphi_1\sin\theta,\\
Y_{\mathrm{JL}}&=\sqrt{R_{\mathrm{JL}}^2+a^2}\sin\varphi_1\sin\theta,\\
Z_{\mathrm{JL}}&=R_{\mathrm{JL}}\cos\theta,
\end{aligned}
\right.\\
&\varphi_1=\varphi+\int^{r_{\mathrm{BL}}} \dfrac{a\,\mathrm{d}s}{s^2-2m s+a^2}+\arctan(\dfrac{a}{R_{\mathrm{JL}}}),\label{eq5}\\
&R_{\mathrm{JL}}=r_{\mathrm{BL}}-m,
\end{align}
which makes the coordinate system $\{X^0_{\mathrm{JL}}=cT_{\mathrm{JL}},X_{\mathrm{JL}},Y_{\mathrm{JL}},Z_{\mathrm{JL}}\}$ harmonic. In this coordinate system the Kerr metric takes the form~\cite{Jiang.2014.GRG}
\begin{widetext}
\begin{align}
\mathrm{d}s^{2}_{\mathrm{JL}}=& -(\mathrm{d}X^{0}_{\mathrm{JL}})^{2} + \frac{R_{\mathrm{JL}}^{2}(R_{\mathrm{JL}}+m)^{2} + a^{2}Z_{\mathrm{JL}}^{2}}{\bigl(R_{\mathrm{JL}}^{2} + \frac{a^{2}}{R_{\mathrm{JL}}^{2}}Z_{\mathrm{JL}}^{2}\bigr)^{2}} \Bigl[ \frac{\bigl( \bm{X}_{\mathrm{JL}} \cdot \mathrm{d}\bm{X}_{\mathrm{JL}} + \frac{a^{2}}{R_{\mathrm{JL}}^{2}}Z_{\mathrm{JL}}\mathrm{d}Z_{\mathrm{JL}} \bigr)^{2}}{R_{\mathrm{JL}}^{2} + a^{2} - m^{2}} + \frac{Z_{\mathrm{JL}}^{2}}{R_{\mathrm{JL}}^{2}} \frac{\bigl( \bm{X}_{\mathrm{JL}} \cdot \mathrm{d}\bm{X}_{\mathrm{JL}} - \frac{R_{\mathrm{JL}}^{2}}{Z_{\mathrm{JL}}}\mathrm{d}Z_{\mathrm{JL}} \bigr)^{2}}{R_{\mathrm{JL}}^{2} - Z_{\mathrm{JL}}^{2}} \Bigr] \notag\\
&+ \frac{2m(R_{\mathrm{JL}}+m)}{(R_{\mathrm{JL}}+m)^{2} + \frac{a^{2}}{R_{\mathrm{JL}}^{2}}Z_{\mathrm{JL}}^{2}} \Bigl[ \frac{R_{\mathrm{JL}}m^{2}a^{2}\bigl(R_{\mathrm{JL}}^{2} - Z_{\mathrm{JL}}^{2}\bigr)\bigl( \bm{X}_{\mathrm{JL}} \cdot \mathrm{d}\bm{X}_{\mathrm{JL}} + \frac{a^{2}}{R_{\mathrm{JL}}^{2}}Z_{\mathrm{JL}}\mathrm{d}Z_{\mathrm{JL}} \bigr)}{\bigl(R_{\mathrm{JL}}^{2} + a^{2} - m^{2}\bigr)\bigl(R_{\mathrm{JL}}^{2} + a^{2}\bigr)\bigl(R_{\mathrm{JL}}^{4} + a^{2}Z_{\mathrm{JL}}^{2}\bigr)} + \frac{a\bigl(Y_{\mathrm{JL}}\mathrm{d}X_{\mathrm{JL}} - X_{\mathrm{JL}}\mathrm{d}Y_{\mathrm{JL}}\bigr)}{R_{\mathrm{JL}}^{2} + a^{2}} + \mathrm{d}X^{0}_{\mathrm{JL}} \Bigr]^{2} \notag\\
&+ \frac{(R_{\mathrm{JL}}+m)^{2} + a^{2}}{R_{\mathrm{JL}}^{2} - Z_{\mathrm{JL}}^{2}} \Bigl[ \frac{R_{\mathrm{JL}}^{2}m^{2}a\bigl(R_{\mathrm{JL}}^{2} - Z_{\mathrm{JL}}^{2}\bigr)\bigl( \bm{X}_{\mathrm{JL}} \cdot \mathrm{d}\bm{X}_{\mathrm{JL}} + \frac{a^{2}}{R_{\mathrm{JL}}^{2}}Z_{\mathrm{JL}}\mathrm{d}Z_{\mathrm{JL}} \bigr)}{\bigl(R_{\mathrm{JL}}^{2} + a^{2} - m^{2}\bigr)\bigl(R_{\mathrm{JL}}^{2} + a^{2}\bigr)\bigl(R_{\mathrm{JL}}^{4} + a^{2}Z_{\mathrm{JL}}^{2}\bigr)} + \frac{R_{\mathrm{JL}}\bigl(Y_{\mathrm{JL}}\mathrm{d}X_{\mathrm{JL}} - X_{\mathrm{JL}}\mathrm{d}Y_{\mathrm{JL}}\bigr)}{R_{\mathrm{JL}}^{2} + a^{2}} \Bigr]^{2},\label{ds2CW}
\end{align}\end{widetext}
where $m=GM/c^2$ and $\bm{X}_{\mathrm{JL}}\cdot \mathrm{d}\bm{X}_{\mathrm{JL}}=X_{\mathrm{JL}}\mathrm{d}X_{\mathrm{JL}}+Y_{\mathrm{JL}}\mathrm{d}Y_{\mathrm{JL}}+Z_{\mathrm{JL}}\mathrm{d}Z_{\mathrm{JL}}$.  The function $R_{\mathrm{JL}}(X_{\mathrm{JL}},Y_{\mathrm{JL}},Z_{\mathrm{JL}})$ is defined by the same implicit or explicit equation as $R(X,Y,Z)$ in Eq.~(\ref{hcan(1)}).

In fact the coordinate transformation \eqref{Jacobi} was already derived by \cite{Cook.1997}, and the corresponding metric was computed to 3PN order in \cite{Will.2017}. Since Jiang and Lin \cite{Jiang.2014.GRG} provided the exact metric \eqref{ds2CW} corresponding to this coordinate transformation, we refer to this coordinate system as Jiang--Lin.

From $\mathrm{d}s^2_{\mathrm{JL}}$ the metric components $g_{\alpha\beta}^{\mathrm{JL}}$, $g^{\alpha\beta}_{\mathrm{JL}}$ and $h^{\alpha\beta}_{\mathrm{JL}}$ can be computed, as shown in the supplementary Mathematica notebook. We expand $h^{\alpha\beta}_{\mathrm{JL}}$ in powers of $G$ and obtain its first order coefficient (the subscript ${\mathrm{JL}}$ in the right-hand side is omitted)
\begin{subequations}
\begin{align}
h^{\alpha\beta}_{\mathrm{JL}}&=Gh^{\alpha\beta}_{\mathrm{JL}[1]}+G^2h^{\alpha\beta}_{\mathrm{JL}[2]}+O(G^3),\\
h_{\mathrm{JL}[1]}^{00} &= -\frac{4MR^3}{c^2(R^4 + a^2Z^2)}, \\
h_{\mathrm{JL}[1]}^{01} &= \frac{2aMR^3Y}{c^2(R^2 + a^2)(R^4 + a^2Z^2)}, \\
h_{\mathrm{JL}[1]}^{02} &= -\frac{2aMR^3X}{c^2(R^2 + a^2)(R^4 + a^2Z^2)}, \\
h_{\mathrm{JL}[1]}^{03} &=h_{\mathrm{JL}[1]}^{ij}= 0,
\end{align}
\end{subequations}
and the second order coefficient
\begin{subequations}
\begin{align}
    h_{\mathrm{JL}[2]}^{00} &= -\frac{M^2 R^2 \bigl(3 a^2+7 R^2\bigr)}{c^4\bigl(a^2+R^2\bigr) \bigl(a^2 Z^2+R^4\bigr)}, \\
    h_{\mathrm{JL}[2]}^{01} &= \frac{2 a M^2 R^2 Y}{c^4\bigl(a^2+R^2\bigr) \bigl(a^2 Z^2+R^4\bigr)}, \\
    h_{\mathrm{JL}[2]}^{02} &= -\frac{2 a M^2 R^2 X}{c^4\bigl(a^2+R^2\bigr) \bigl(a^2 Z^2+R^4\bigr)}, \\
    h_{\mathrm{JL}[2]}^{03} &= 0, \\
    h_{\mathrm{JL}[2]}^{11}&=-\frac{M^2 \bigl(R^2 X^2 \bigl(R^2-a^2\bigr)+a^2 \bigl(a^2+R^2\bigr) \bigl(R^2-Z^2\bigr)\bigr)}{c^4\bigl(a^2+R^2\bigr)^2 \bigl(a^2 Z^2+R^4\bigr)}\notag\\
    &\quad-\frac{2M^2 a R^3 X Y}{c^4\bigl(a^2+R^2\bigr)^2 \bigl(a^2 Z^2+R^4\bigr)},\\
    h_{\mathrm{JL}[2]}^{22}&=-\frac{M^2 \bigl(R^2 Y^2 \bigl(R^2-a^2\bigr)+a^2 \bigl(a^2+R^2\bigr) \bigl(R^2-Z^2\bigr)\bigr)}{c^4\bigl(a^2+R^2\bigr)^2 \bigl(a^2 Z^2+R^4\bigr)}\notag\\
    &\quad+\frac{2M^2 a R^3 X Y}{c^4\bigl(a^2+R^2\bigr)^2 \bigl(a^2 Z^2+R^4\bigr)},\\
    h_{\mathrm{JL}[2]}^{33}&=-\frac{M^2 Z^2}{c^4(a^2 Z^2+R^4)},\\
    h_{\mathrm{JL}[2]}^{12}&=\frac{-M^2 R^2 X Y \bigl(R^2-a^2\bigr)}{c^4\bigl(a^2+R^2\bigr)^2 \bigl(a^2 Z^2+R^4\bigr)}\notag\\
    &\quad+\frac{a M^2 R^3 \bigl(X^2-Y^2\bigr)}{c^4\bigl(a^2+R^2\bigr)^2 \bigl(a^2 Z^2+R^4\bigr)},\\
    h_{\mathrm{JL}[2]}^{23}&=\frac{-M^2 R^2 YZ}{c^4\bigl(a^2+R^2\bigr) \bigl(a^2 Z^2+R^4\bigr)}\notag\\
    &\quad+\frac{a M^2 R XZ}{c^4\bigl(a^2+R^2\bigr) \bigl(a^2 Z^2+R^4\bigr)},\\
    h_{\mathrm{JL}[2]}^{31}&=\frac{-M^2 R^2 X Z}{c^4\left(a^2+R^2\right) \left(a^2 Z^2+R^4\right)}\notag\\
    &\quad-\frac{a M^2 R Y Z}{c^4\left(a^2+R^2\right) \left(a^2 Z^2+R^4\right)}.
\end{align}
\end{subequations}

Expanding the left-hand side of \eqref{proposition} to 2PM and substituting the JL metric to the right-hand side, we get
\begin{subequations}
\begin{align}
&h_{(1)}^{\alpha\beta}[\mathrm{I}_L^0,\mathrm{J}_L^0,\mathrm{W}_L^0,\mathrm{X}_L^0,\mathrm{Y}_L^0,\mathrm{Z}_L^0]\overset{\mathrm{s}}{=} h_{\mathrm{JL}[1]}^{\alpha\beta},\\
&h_{(1)}^{\alpha\beta}[\mathrm{I}_L^1,\mathrm{J}_L^1,\mathrm{W}_L^1,\mathrm{X}_L^1,\mathrm{Y}_L^1,\mathrm{Z}_L^1]\notag\\
+&h_{(2)}^{\alpha\beta}[\mathrm{I}_L^0,\mathrm{J}_L^0,\mathrm{W}_L^0,\mathrm{X}_L^0,\mathrm{Y}_L^0,\mathrm{Z}_L^0]\overset{\mathrm{s}}{=} h_{\mathrm{JL}[2]}^{\alpha\beta}.
\end{align}
\end{subequations}
Since $h^{\alpha\beta}_{\mathrm{JL}[1]}$ coincides with the closed form of \eqref{hcan(1)}, we can immediately read off the 0PM source moments
\begin{subequations}
\begin{align}
    \mathrm{I}_L^0&=\left\{\begin{aligned}
        &(-1)^{l/2}M a^l (\bm{e}_z)_{\langle L\rangle},&&\text{$l$ even},\\
        &0,&&\text{$l$ odd},
        \end{aligned}\right.\\
    \mathrm{J}_L^0&=\left\{\begin{aligned}
        &(-1)^{(l-1)/2}\dfrac{l+1}{2l}M c a^l (\bm{e}_z)_{\langle L\rangle},&&\text{$l$ odd},\\
        &0,&&\text{$l$ even}.
        \end{aligned}\right.\\
    \mathrm{W}_L^0&=\mathrm{X}_L^0=\mathrm{Y}_L^0=\mathrm{Z}_L^0=0,
\end{align}
\end{subequations}
Thus we derive
\begin{subequations}
\begin{align}
    h_{(1)}^{00}[\mathrm{A}_L^1]&=h_{(1)}^{0i}[\mathrm{A}_L^1]=0,\\
    h_{(1)}^{11}[\mathrm{A}_L^1]&\overset{\mathrm{s}}{=}-\frac{2M^2 a R^3 X Y}{c^4\bigl(a^2+R^2\bigr)^2 \bigl(a^2 Z^2+R^4\bigr)},\\
    h_{(1)}^{22}[\mathrm{A}_L^1]&\overset{\mathrm{s}}{=}+\frac{2M^2 a R^3 X Y}{c^4\bigl(a^2+R^2\bigr)^2 \bigl(a^2 Z^2+R^4\bigr)},\\
    h_{(1)}^{33}[\mathrm{A}_L^1]&=0,\\
    h_{(1)}^{12}[\mathrm{A}_L^1]&\overset{\mathrm{s}}{=}+\frac{a M^2 R^3 \bigl(X^2-Y^2\bigr)}{c^4\bigl(a^2+R^2\bigr)^2 \bigl(a^2 Z^2+R^4\bigr)},\\
    h_{(1)}^{23}[\mathrm{A}_L^1]&\overset{\mathrm{s}}{=}+\frac{a M^2 R XZ}{c^4\bigl(a^2+R^2\bigr) \bigl(a^2 Z^2+R^4\bigr)},\\
    h_{(1)}^{31}[\mathrm{A}_L^1]&\overset{\mathrm{s}}{=}-\frac{a M^2 R Y Z}{c^4\left(a^2+R^2\right) \left(a^2 Z^2+R^4\right)},
\end{align}
\end{subequations}
and read off the 1PM source moments
\begin{subequations}
\begin{align}
    \mathrm{I}_L^1&=\mathrm{J}_L^1=   \mathrm{W}_L^1=\mathrm{X}_L^1=\mathrm{Y}_L^1=0,\\
    \mathrm{Z}_L^1&=\left\{\begin{aligned}
        &(-1)^{(l-1)/2} \frac{l+1}{4 l (l+2)}M^2 a^l (\bm{e}_z)_{\langle L\rangle},&&\text{$l$ odd},\\
        &0,&&\text{$l$ even}.
        \end{aligned}\right.
\end{align}
\end{subequations}
We thus find that the Kerr metric in Jiang--Lin coordinates possesses non-zero gauge moments $\mathrm{Z}_L=G\mathrm{Z}_L^1+O(G^2)$ at 1PM level. In this sense Jiang--Lin coordinates have more gauge redundancy than canonical coordinates, where all gauge moments vanish at every PM order.

The situation for the harmonic coordinates discussed in \cite{1986GReGr..18..805R,1998ChPhL..15..313L,2020CQGra..37t7002Z} is similar. We do not present the analysis results any further.
\section{Summary and discussion}

In this paper we have constructed the canonical harmonic coordinates to 4PM. For the Schwarzschild case, the whole canonical metric has been constructed to all PM orders.

In order to relate our canonical harmonic coordinates to other existing coordinates, we have computed the Ricci tensor and the Kretschmann scalar $K=R_{\alpha\beta\mu\nu}R^{\alpha\beta\mu\nu}$ of the 4PM canonical metric.
The Ricci tensor vanishes identically at 4PM order, i.e., $R_{\alpha\beta}=O(G^5)$, which directly verifies that the canonical metric satisfies the vacuum Einstein field equations to 4PM order. The Kretschmann scalar of the canonical metric is
\begin{widetext}
\begin{align}
    K&=\frac{G^2 M^2}{c^4}\Bigg(\frac{48  R^6 \left(-a^6 Z^6+15 a^4 R^4 Z^4-15 a^2 R^8 Z^2+R^{12}\right)}{\left(a^2 Z^2+R^4\right)^6}\notag\\
    &\quad-\frac{288 G M R^9 \left(-7 a^6 Z^6+35 a^4 R^4 Z^4-21 a^2 R^8 Z^2+R^{12}\right)}{c^2 \left(a^2 Z^2+R^4\right)^7}\notag\\
    &\quad+\frac{1008 G^2 M^2 R^8 \left(a^8 Z^8-28 a^6 R^4 Z^6+70 a^4 R^8 Z^4-28 a^2 R^{12} Z^2+R^{16}\right)}{c^4 \left(a^2 Z^2+R^4\right)^8}\notag\\
    &\quad-\frac{2688 G^3 M^2 R^{11} \left(9 a^8 Z^8-84 a^6 R^4 Z^6+126 a^4 R^8 Z^4-36 a^2 R^{12} Z^2+R^{16}\right)}{c^{6} \left(a^2 Z^2+R^4\right)^9}+O\left(G^4\right)\Bigg),
\end{align}
\end{widetext}
which agrees with the Kretschmann scalar in Boyer--Lindquist coordinates \citep{Henry.2000} under the following relations
\begin{align}
    r_{\mathrm{BL}}&=R+m+O(m^4),\label{eq2}\\
    \cos\theta&=\dfrac{Z}{R}+O\left(m^4\right),\label{eq3}
\end{align}
This matching of the curvature scalar indicates that $R$ equals $R_{\mathrm{JL}}=r_{\mathrm{BL}}-m$ at least till order $O(m^3)$, or equivalently $O(G^3)$.

Based on $R(X,Y,Z)$ we introduce the auxiliary angular variables $\theta_2$ and $\varphi_2$ by
\begin{align}
&\left\{
\begin{aligned}
T&=t,\\
X&=\sqrt{R^2+a^2}\,\cos\varphi_2\,\sin\theta_2,\\
Y&=\sqrt{R^2+a^2}\,\sin\varphi_2\,\sin\theta_2,\\
Z&=R\cos\theta_2.
\end{aligned}
\right.
\end{align}

Eqs.~(\ref{eq2}) and (\ref{eq3}) guide us to the following coordinate transformation
\begin{align}
    R&=r_{\mathrm{BL}}-m+O(m^4),\\
    \theta_2&=\theta+O(m^4),\\
    \varphi_2&=\varphi+\int^{r_{\mathrm{BL}}} \dfrac{a\,\mathrm{d}s}{s^2-2m s+a^2}+\arctan(\dfrac{a}{R})\notag\\
    &\quad -\frac{m^2}{2 a^2}\left(\frac{a R}{a^2+R^2}-\arctan\left(\frac{a}{R}\right)\right)
    +O(m^4).\label{eq4}
\end{align}
We have verified that the above coordinate transformation can produce the 3PM canonical metric $g^{\alpha\beta}_{\rm can3PM}$ from Boyer-Lindquist metric $g^{\mu\nu}_{\rm BL}$. The 4PM coordinate transformation linking the canonical coordinates to Boyer-Lindquist coordinates remains an unresolved problem demanding further investigation. Comparing the above equation to Eq.~(\ref{eq5}), we find that the canonical angular variable $\varphi_2$ is different from the Jiang--Lin angular variable $\varphi_1$ starting from $O(G^2)$, which corresponds to the difference between $h^{\alpha\beta}_{\rm can(2)}$ and $h^{\alpha\beta}_{\rm JL[2]}$ at the 2PM order.

\acknowledgments
This work is supported by the National Key Research and Development Program of China (Nos. 2021YFC2203001), the National Natural Science Foundation of China (No.~12475049).

\bibliographystyle{unsrt}
\bibliography{ref}
\end{document}